\documentclass[11pt]{article}
\usepackage{moriond}
\usepackage{natbib}
\usepackage{amsmath}
\usepackage{amsfonts}
\usepackage{amssymb}
\newcommand{\unit}[1]{\ensuremath{\;{\mathrm{#1}}}} 

\renewcommand{\deg}{\ensuremath{^\circ}}

\newcommand{\Photo}{}

\usepackage{fancyhdr}
\begin{document}
\vspace*{2cm}
\title{Search for High Energy GRB Neutrino Emission with ANTARES}

\author{J. Schmid on behalf of the ANTARES collaboration}

\address{Erlangen Centre for Astroparticle Physics --- Erwin-Rommel-Str.1 --- 91058 Erlangen\\ Julia.Schmid@fau.de}

\maketitle\abstracts{ANTARES is the largest high-energy neutrino telescope in the Northern Hemisphere. 
A search for neutrinos in coincidence with gamma-ray bursts using ANTARES data from late 2007 to 2011 is presented here. An extended maximum likelihood ratio search was employed to optimise the discovery potential for a neutrino signal as predicted by the numerical NeuCosmA model. No significant excess was found, so 90\% confidence upper limits on the fluxes as expected from analytically approximated neutrino-emission models as well as on up-to-date numerical predictions were placed.}

\section{Introduction}
The ANTARES neutrino telescope \citep{Antares11a} is located in the Mediterranean Sea at a depth of $2.4\unit{km}$. The detector consists of 12 vertical strings, separated from each other by a typical distance of $70\unit{m}$. Each string is anchored to the seabed and held upright by a buoy at the top; 
over a length of $350\unit{m}$, it is equipped with 25 triplets of photo-multiplier tubes (PMTs), building a 3-dimensional array of 885 PMTs in total. 
The instrumented volume is $\sim  \! 0.02\unit{km}^3$.

The main scientific purpose of ANTARES is the search for astrophysical neutrinos, which are detected via their charged-current interactions in the Earth and the subsequent Cherenkov emission of secondary charged leptons in the water of the Mediterranean Sea. Gamma-ray bursts (GRBs) are suitable cosmic candidate sources for neutrino telescopes, as they are thought to accelerate not only electrons --- leading to the observed gamma rays --- but also protons, which would yield the emission of high-energy neutrinos. 

In the prevailing fireball model for gamma-ray bursts as proposed for example by M{\'e}sz{\'a}ros and Rees \citep{Meszaros93a}, the observed electromagnetic radiation is explained by synchrotron radiation and subsequent inverse Compton scattering of relativistic shock-accelerated electrons \citep{Rees92a}. 
If protons are also accelerated in the shock outflow, high-energy neutrinos would accompany the electromagnetic signal of the burst \citep{Waxman97a}.
 The detection of GRB neutrinos would be unambiguous proof for hadronic acceleration in cosmic sources and could also serve to explain the origin of the  cosmic-ray flux at ultra-high energies. 

For the analysis presented in the following, we utilise ANTARES data from the end of 2007 to 2011. % after completion of the detector.
The total integrated live-time of the data in coincidence with the selected sample of 296 GRBs is 6.6 hours. 
For a more detailed description of the analysis scheme, see \citep{Antares13b}.

\section{Analysis}
\subsection{GRB Parameters and Selection}
The parameters needed for the search and the simulation of expected neutrino fluxes are primarily obtained from different gamma-ray-burst catalogues provided by \textsl{Swift} \citep{Sakamoto11a} and the \textsl{Fermi} GBM \citep{Goldstein12a}. 
The catalogues were then completed using a table supplied by the IceCube collaboration \citep{IceCube11a}, which is created by parsing the Gamma-ray Coordinates Network notices\footnote{GCN: \url{gcn.gsfc.nasa.gov/gcn3_archive.html}}.
In case a parameter could not be measured, standard values as given in \citep{IceCube11a} are used to calculate the spectra. 

\begin{figure}[h!]
   	\centering
	    \includegraphics[height=0.17\textheight]{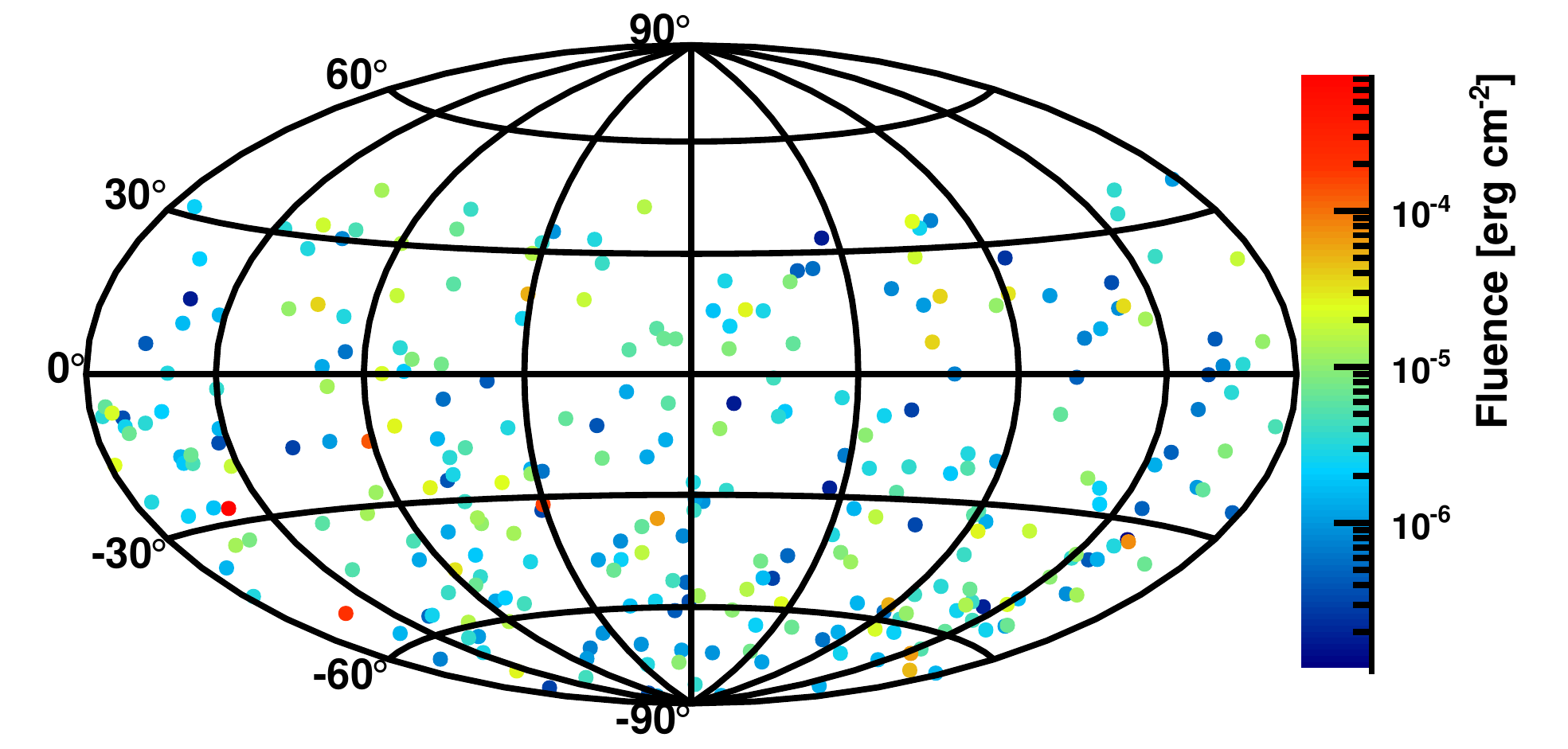}
	    \includegraphics[height=0.17\textheight]{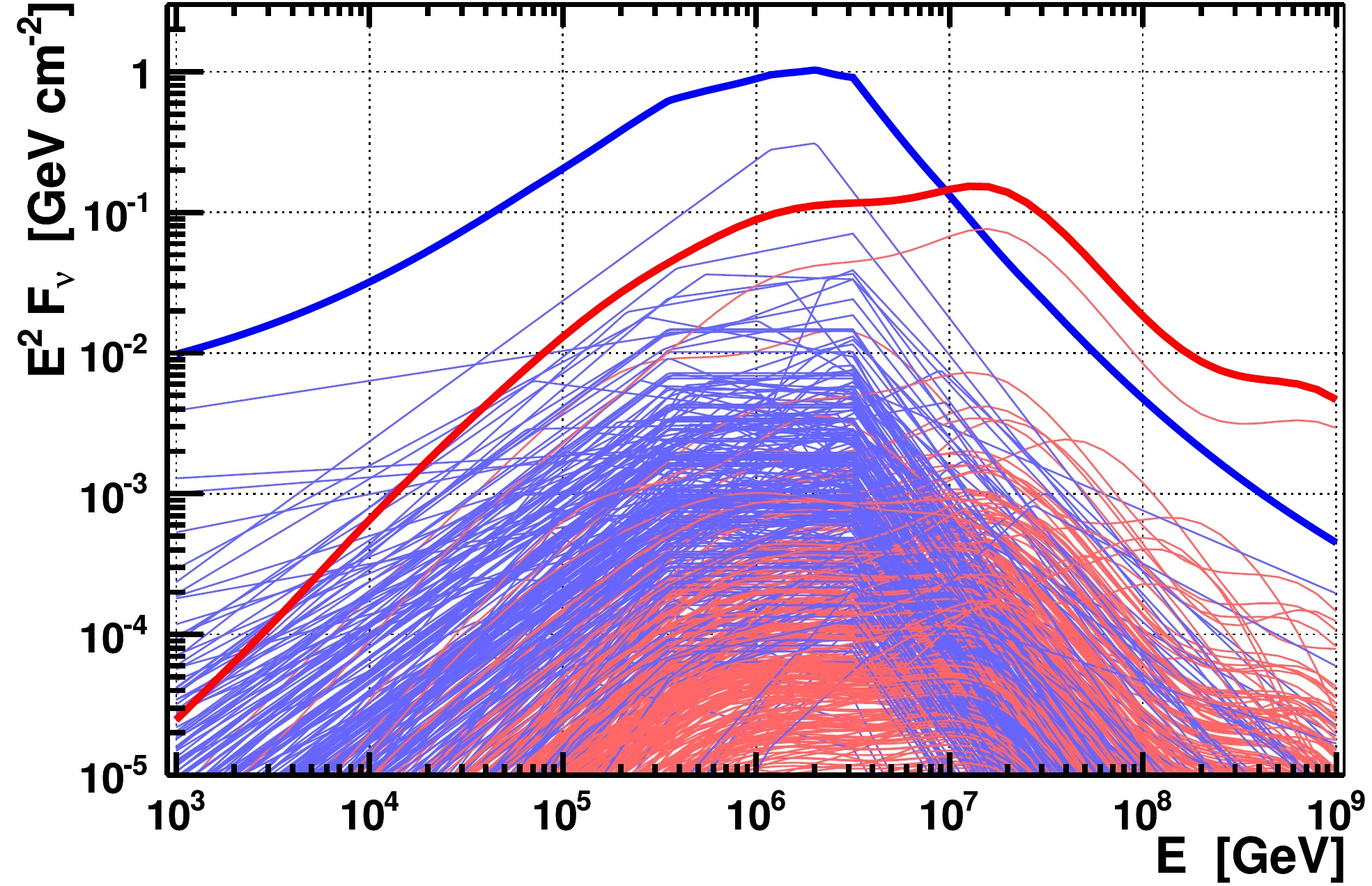}
      \caption{\emph{Left:} Sky distribution of the selected 296 gamma-ray bursts in equatorial coordinates. The photon fluence of each burst is indicated by the colours. The ANTARES detector is sensitive to both the Southern and Northern Hemisphere up to a declination of $47\deg$, with an instantaneous field of view of $2\pi$. \emph{Right:}\label{fig:spectra} Individual neutrino spectra of the 296 GRBs. Blue lines show the expectations due to the analytical model \citep{Guetta04a}, whereas red lines show the NeuCosmA predictions \citep{Huemmer12a}, respectively. Thick lines show the sum of all individual spectra. }
      \label{fig:grb_coordinates}
    \end{figure}

For the final sample, gamma-ray bursts are excluded when neither spectral nor fluence information was available and when no duration was given.
Short GRBs are also discarded since this class is much less understood. 
In addition we require both that the GRBs were located below the local horizon for ANTARES and that the detector was taking reliable physics data during the burst. 
In total, 296 bursts were selected. 
Their distribution on the sky is shown in Fig.~\ref{fig:grb_coordinates}.

\subsection{Gamma-Ray Burst Neutrino Emission}
The expected neutrino fluxes from gamma-ray bursts are usually derived from the spectrum of the Fermi-accelerated protons and the measured photon spectrum.
 Different models have been developed to calculate the neutrino flux from photohadronic interactions; up-to-date models like NeuCosmA \citep[see Ref.][and references therein]{Huemmer12a} feature detailed numerical simulations of the underlying processes.
 % including up-to-date fully numerical simulations of the processes, like the NeuCosmA model \citep[as described in detail in][]{Huemmer10a,Huemmer12a,Baerwald12b}. 
 The analysis presented here relies on the predictions made by this model. 
Figure~\ref{fig:spectra} shows the spectra for the numerical NeuCosmA model in comparison to the widely-used analytical approximation \citep{Guetta04a}, which overestimates for instance the pion production efficiency \citep{Huemmer12a}.

\subsection{Signal and Background Probability Density Functions}
For each gamma-ray burst, signal neutrino events according to the expected NeuCosmA fluxes are simulated with high statistics and then reconstructed in order to compute the detector's acceptance and the spread of events around the actual burst directions.
This distribution yields the signal probability density function (PDF) labelled $\mathcal{S}(\delta)$, where $\delta$ represents the space angle between the reconstructed event direction and the GRB's coordinates. 

The background PDF $\mathcal{B}(\delta)$ is considered to be flat within a $10\deg$ search cone around each burst position.
In order to estimate the expected mean number of background events $\mu_\mathrm{b}$ for each burst as realistically as possible, real reconstructed data events are used.
However, as the number of upgoing events is very low ($\sim 4$ per day, see \citep{Antares12c}), large time periods are needed to yield enough statistics, which in turn requires averaging over different detector conditions (in particular due to seasonal variations of the optical background). 
To compensate for this, we first estimate the average reconstructed event rate in the GRB's direction using data from the whole  late-2007 to 2011 period, then adjust it for varying detector conditions.

The reconstruction algorithm returns the track fit quality parameter $\Lambda$; 
a cut on this parameter selects well-reconstructed events and is later-on used to optimise the analysis.
Both $\mathcal{S}(\delta)$ and $\mathcal{B}(\delta)$ depend on the final choice of the cut on $\Lambda$. 

\subsection{Search Optimisation}
Pseudo-experiments are generated that randomly draw signal and background events $i$ with space angle $\delta_i$ from the normalised PDFs $\mathcal{B}(\delta)$ and $\mathcal{S}(\delta)$ corresponding to each $\Lambda$ cut.
For each pseudo-experiment with $n_\mathrm{tot}$ events, the test statistic $Q$ is calculated: 
    \begin{equation}
      Q = \max_{\mu^{\prime}_\mathrm{s} \in [0,n_\mathrm{tot}]} \left( \sum\limits_{i=1}^{n_\text{tot}} \log \frac{\mu^\prime_\mathrm{s} \cdot \mathcal{S}(\delta_i) + {\mu}_\mathrm{b} \cdot \mathcal{B}(\delta_i)}{{\mu}_\mathrm{b} \cdot \mathcal{B}(\delta_i)}  - (\mu^{\prime}_\mathrm{s} +{\mu}_\mathrm{b}) \right).
      \label{eq:ex_max_likelihood}
    \end{equation}
This is the so-called Extended Maximum Likelihood Ratio \citep{Barlow90a} with an a-priori knowledge of the expected number of background events ${\mu}_\mathrm{b}$.
Larger values of $Q$ indicate that the measurement is more compatible with the signal hypothesis.

The distributions of the test statistic for different numbers of injected events are used to evaluate the model discovery potential $\mathcal{MDP}$ for a given number of expected signal events $\mu_\mathrm{s}$ as predicted by the NeuCosmA model. 
The cut on the quality parameter $\Lambda$ is then chosen as that which maximises the $\mathcal{MDP}$. 
Figure~\ref{fig:pe_mdp} shows the $\mathcal{MDP}(\mu_\mathrm{s})$ using PDFs of GRB110918 for $3\sigma$, $4\sigma$ and $5\sigma$ for an arbitrary number of expected signal events.

\section{Results}
Using the strategy outlined above, we analysed ANTARES data from the end of 2007 to 2011 searching for neutrino events in coincidence with the selected gamma-ray bursts and within $10\deg$ around each.
No data events passed the event selection cuts within the accumulated search duration of 6.55 hours.
Hence, the measured $Q$-value is zero. 

\begin{figure}[h!] \centering
 		\includegraphics[height = 0.17\textheight]{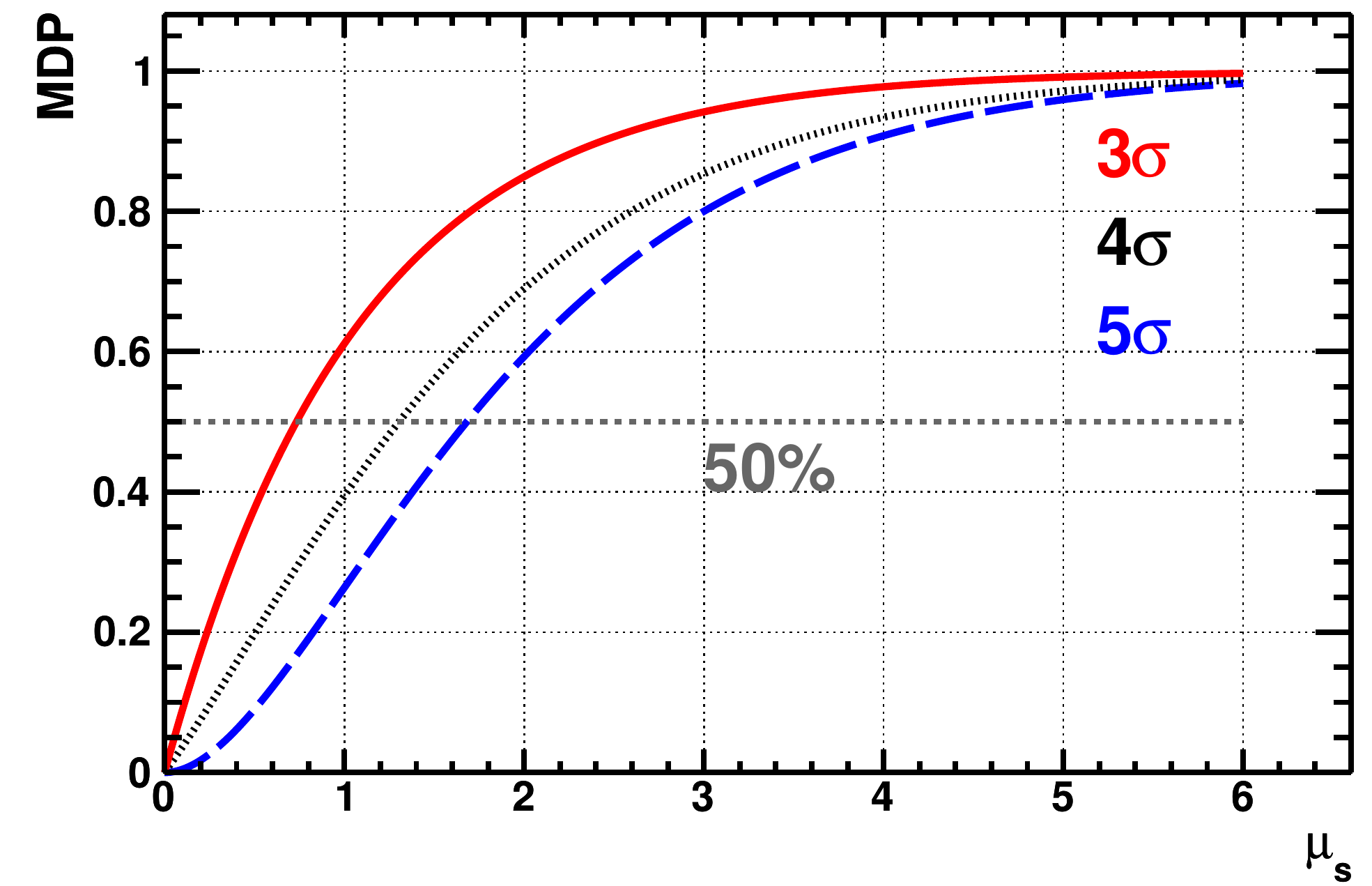}	
		\includegraphics[height = 0.17\textheight]{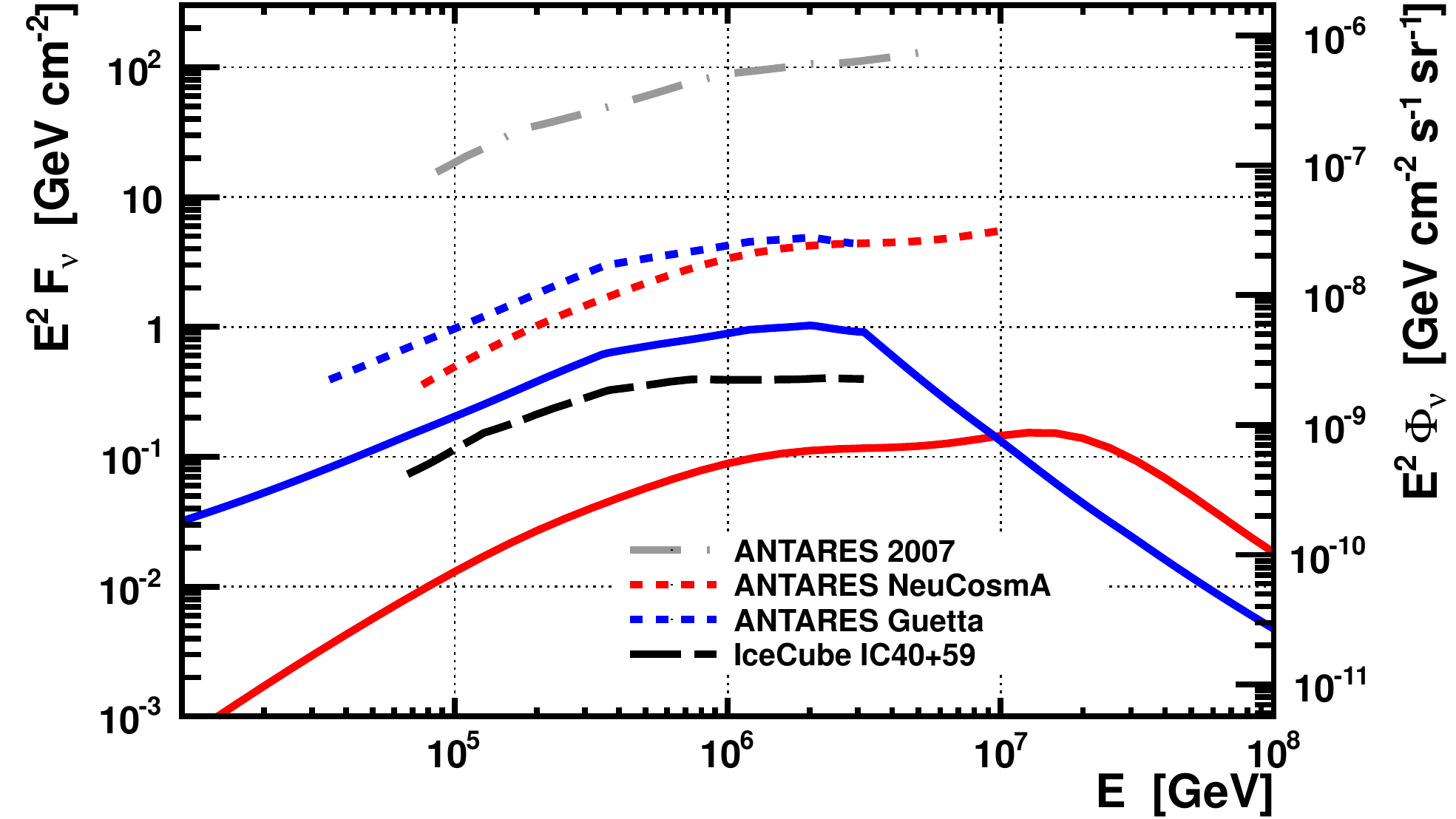}
\caption{\emph{Left:}\label{fig:pe_mdp} Model Discovery Potential versus any mean number of expected signal events $\mu_\mathrm{s}$ for $3\sigma$ (red solid line), $4\sigma$ (black dotted) and $5\sigma$ significance (blue dashed), with PDFs for GRB110918 and $\Lambda > -5.5$.
\label{fig:limit} 
\emph{Right:} Solid lines: sum of the 296 individual gamma-ray burst neutrino spectra used in this analysis, for the analytical model in blue and the NeuCosmA model in red. Dashed lines show the limits on these fluxes.
The IceCube IC40+59 limit \citep[][]{IceCube12a} on the neutrino emission from 300 GRBs
and the first ANTARES limit from its construction phase in 2007 using 40 GRBs \citep{Antares13a} are also shown in black (dashed) and grey (dash-dotted), respectively.}
\end{figure}

In total, 0.06 signal events were predicted from the NeuCosmA model (0.5 from the analytical model) against a background of 0.05 events.
The 90\% C.L. upper limit on the signal can be set to 2.3 events,
thus limits can be placed on the cumulative flux of the whole sample as shown in Fig.~\ref{fig:limit}. 

Limits from previous analyses are also shown: 
the ANTARES limit \citep{Antares13a} from the construction phase of the detector in 2007 as well as the IceCube limit \citep[][]{IceCube12a} using the IC40 and IC59 detector configurations used the analytical model predictions.
The right-hand axis of  Fig.~\ref{fig:limit} (b) shows how the limits on the individual samples translate into a limit on the inferred quasi-diffuse flux, 
assuming that the selections represent average burst distributions and that the annual rate of long bursts is 667 per year.

We have presented the first analysis that relies on up-to-date numerical simulations of neutrino emission from GRBs. 
It has been shown that the expected fluxes are one order of magnitude lower than predicted by prevailing analytical approaches \citep[e.g.][]{Guetta04a}. 
Hence, existing limits have not put any constraints on realistic neutrino emission models; i.e. the fireball paradigm has not yet been probed by neutrino telescopes.
Nevertheless, the collection of more and more data with active experiments as well as planned neutrino telescopes like KM3NeT will certainly allow the widely-established fireball paradigm to be probed in the near future.

\section*{Acknowledgments}
This work was partly funded by the Studienstiftung des Deutschen Volkes.

\end{document}